# The Case for Non-Cryogenic Comet Nucleus Sample Return

*A White Paper submitted to the Planetary Science Decadal Survey 2023-2032 reflecting the viewpoints of three New Frontiers comet sample return missions proposal teams, CAESAR, CONDOR, and CORSAIR.*

Keiko Nakamura-Messenger* (JSC), Alexander G. Hayes (Cornell University), Scott Sandford (ARC), Carol Raymond (JPL), Steven W. Squyres (Cornell University), Larry R. Nittler (Carnegie), Samuel Birch (MIT), Denis Bodewits (Auburn University), Nancy Chabot (JHU/APL), Meenakshi Wadhwa (ASU), Mathieu Choukroun (JPL), Simon J. Clemett (JSC), Maitrayee Bose (ASU), Neil Dello Russo (JHU/APL), Jason P. Dworkin (GSFC), Jamie E. Elsila (GSFC), Kenton Fisher (JSC), Perry Gerakines (GSFC), Daniel P. Glavin (GSFC), Julie Mitchell (JSC), Michael Mumma (GSFC), Ann. N. Nguyen (JSC), Lisa Pace (JSC), Jason Soderblom (MIT), Jessica M. Sunshine (University of Maryland)  *keiko.nakamura-1@nasa.gov

**Summary:**
Comets hold answers to mysteries of the Solar System by recording presolar history, the initial states of planet formation and prebiotic organics and volatiles to the early Earth. Analysis of returned samples from a comet nucleus will provide unparalleled knowledge about the Solar System's starting materials and how they came together to form planets and give rise to life:

1. *How did comets form?*
2. *Is comet material primordial, or has it undergone a complex alteration history?*
3. *Does aqueous alteration occur in comets?*
4. *What is the composition of cometary organics?*
5. *Did comets supply a substantial fraction of Earth's volatiles?*
6. *Did cometary organics contribute to the homochirality in life on Earth?*
7. *How do complex organic molecules form and evolve in interstellar, nebular, and planetary environments?*
8. *What can comets tell us about the mixing of materials in the protosolar nebula?*

## 1.  Introduction

Comets are time capsules from the birth of our Solar System that record presolar history and the initial states of planet formation, and likely contributed prebiotic organics and volatiles to the early Earth. Analysis of samples from a comet nucleus can provide unparalleled knowledge about the nature of Solar System starting materials and how these components came together to form planets and play a role in the origin of life [14]. Laboratory examination of comet nucleus samples will also provide ground truth for remote sensing observations of innumerable icy bodies in the outer Solar System [15]. In general, missions that return planetary samples enable scientific discoveries far surpassing those that can be achieved in situ (see Figure 1 and [16]). For comet nucleus samples, the precision and sensitivity of Earth-based laboratory analyses will lead to major advances in disciplines ranging from molecular cloud chemistry to the evolution of the Solar System's volatile reservoirs, including the first steps of prebiotic organic synthesis.

Rosetta's investigation of comet 67P/Churyumov-Gerasimenko revealed extraordinary new details on cometary geology and surface processes, but these findings leave open fundamental questions regarding cometary origins [17]. Instruments on the Rosetta orbiter determined that 67P's coma dust contained as much as 50% organic matter [24] and in-situ analyses gave tantalizing hints of the presence of complex molecules. Unfortunately, instrument limitations and an anomalous touchdown by the Philae lander left the details of cometary organic chemistry undetermined [25]. This leaves open the question of whether or not comets contain prebiotic organic compounds that may have contributed to the origin of life on Earth.

The key to deciphering the origins of comets lies in the chemical and isotopic makeup of their volatiles. Moreover, the compositions of comet volatiles hold clues to the origins of Earth's water and organic matter [26-27]. Non-cryogenic comet sample return can achieve many of these priority science objectives, and act as a pathfinder for eventual but far more ambitious and costly cryogenic sample return [2; 29-31]. Acquiring a cryogenic sample at temperatures < 90 K and maintaining its integrity and temperature during return to Earth, represent major technological challenges that will require substantial investment [34]. Indeed, a cryogenic sample return mission would likely need to be a flagship with a multi-billion-dollar budget. Non-cryogenic comet nucleus sample return therefore represents a logical next step in comet science, and will reveal the composition of comet nuclei, the physical process of accretion and formation, and help constrain the origin of Earth's volatiles and prebiotic organic inventory [2].

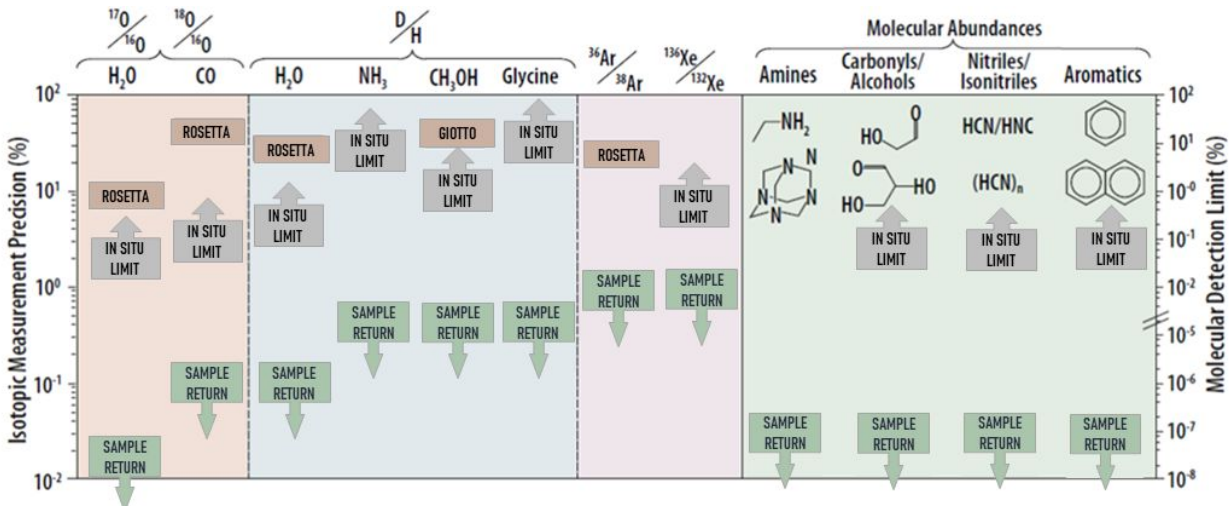

Figure 1: Examples of volatile measurement capabilities in returns sampled vs. in situ spacecraft. Left: Isotopic measurement precision of selected volatiles associated with known cometary ices. Right: Molecular detection limits for astrobiologically significant classes of organics. Downward arrows reflect expected improvements in terrestrial measurement capabilities in the coming decades. Data from [1-13].

## 2. Lessons Learned Since Visions and Voyages

At the time the last Decadal Survey (*Visions and Voyages*) [35] was conducted, five comets (1P/Halley, 19P/Borrelly, 81P/Wild 2, 9P/Tempel 1, and 103P/Hartley 2) had been visited by spacecraft flybys by the Soviet Union (VEGA 1/2), ESA (Giotto), and NASA (Deep Space 1, Stardust / Stardust-NExT, and Deep Impact / EPOXI). From these missions, we learned that comet nuclei are dominated by refractories (i.e., an icy dirtball rather than a dirty snowball) [36], are dark, porous, and low density [e.g., 37], have rough surfaces and are morphologically diverse [38-40], and contain abundant hypervolatile species (e.g., CO, $N_2$) that condensed below 30 K [41]. The spacecraft missions conducted to date give conflicting views, however, of comet formation (e.g., [18] vs [19]), requiring sample return to determine whether comets contain large fractions of interstellar materials [20-21] or are dominated by materials processed in the protoplanetary disk [22; 23]. Stardust was the first mission to bring cometary material back to Earth, and the analysis of the minute amounts of refractory material (< $10^{-6}$ gram) returned from the coma of 81P/Wild 2 spurred a flurry of analytical papers on the sample's mineralogy and petrology [23; 42], isotopic and elemental compositions [43; 44], and organics [45]. The large mineralogic diversity observed in the sample and the detection of high-temperature minerals suggested large-scale mixing between the inner and outermost parts of the solar nebula [23; 46]. These findings left open the question of whether the observed diversity among comets represented differences in their formation environment or later evolution. This question was to be addressed in part by ESA's Rosetta mission [47].

Rosetta launched in 2004, reached comet 67P/Churyumov-Gerasimenko (67P) in August 2014 [48], and followed the development of activity on 67P through perihelion, ceasing flight operations in September 2016. Images with decimeter-scale resolution demonstrated that 67P's geology is extremely diverse, with all morphologies observed in other Jupiter Family Comets (JFCs) being present on 67P [50; 51]. These similarities suggest that JFCs are similar, and that observed differences reflect different states along a common evolutionary pathway [47; 50], with additional influences due to nucleus shape, rotation, obliquity, and dynamical history [55]. Water ice patches were observed both as morning frosts and at the base of cliffs and along the margins of exposed depressions [57-58]. Activity and surface changes were observed to occur across all parts of the nucleus [63-65], suggesting that water ice is universally present at shallow depths (< 25 cm) [63-64].

Findings from Rosetta's ROSINA instrument confirmed that comet formation took place at very low temperatures (< 25 K) and that much of its volatile makeup could be of interstellar origin [18; 66]. Remote sensing in the near-IR [67], ROSINA mass spectrometer measurements [68], and direct sampling by the Philae lander [69] demonstrated that 67P's surface contains refractory and semi-volatile organics, as well as both volatile and hypervolatile ices. Hypervolatile species, including CO and $CO_2$, were present in abundances that exceed terrestrial detection limits by more than five orders of magnitude [68]. There remains considerable debate on the total amount of volatile materials, however, as uncertainties associated with the derivation of the refractory-to-ice mass ratio from Rosetta observations make it difficult to constrain.

Rosetta's ROSINA, Ptolemy, and COSAC instruments detected most known comet volatiles, as well as new volatiles including P (likely from $PH_3$), large quantities of $O_2$, and $N_2$ [10; 69; 71]. Philae's COSAC instrument also identified 16 semi-volatile organic species, including many N-bearing species and four compounds (acetone, acetamide, methyl, and ethyl isocyanate) not previously identified on comets [10]. Finally, and perhaps most surprisingly, Rosetta found that refractory material ("dust") is 45 weight % organic, confirming that comets are massive reservoirs of organic materials [72]. Together, the close-up views of comets revealed by Stardust and Rosetta make it clear that comet surface samples contain the mineralogical, molecular, elemental, and isotopic evidence needed to test formation hypotheses and investigate prebiotic pathways via laboratory analysis of returned samples. However, in part because most of the planned experiments on Philae were not conducted on 67P, there remain many open questions about the nature and chemical composition of the non-volatile organic component of comets, including the presence and chirality of prebiotic organic compounds including amino acids and sugars. At 67P, the Rosetta mass spectrometer ROSINA identified the amino acid glycine and two volatile amines, methylamine and ethylamine [73], however these molecules are all achiral. Furthermore, the wet chemistry experiments planned for the COSAC instrument on Philae that could have detected chiral molecules were not conducted. Therefore, the composition of chiral amino acids and the presence of other important prebiotic organic molecules in comets that could have contributed to the origin of life on Earth remain largely unknown.

## 3. Science Overview

While multiple sampling mechanisms and mission concepts have been proposed for non-cryogenic comet nucleus sample return [29-31], all are designed to acquire and return ~100 g of refractory comet material to Earth for laboratory analysis. Volatile gasses can either be analyzed in situ [e.g., 30] or brought back to Earth for laboratory study [e.g., 29]. In either case, analyses of both volatile and non-volatile samples are needed to discern the origin, history, and composition of Solar System starting materials and the nature of volatiles and organics contributed by comets to the early Earth. While a thorough description of the analysis techniques and scientific hypotheses that can be tested by studying a returned comet sample is outside the scope of this white paper, below we provide a brief overview of some of the most compelling questions that can be addressed. In fact, a complete description is impossible. One of the greatest benefits of sample return, as shown by recent studies of Apollo lunar samples [e.g., 74], is that the mission's most important instrumentation is in terrestrial laboratories rather than fixed at launch. It is inevitable that some of the most important questions about comet nucleus samples will be answered with next-generation laboratory instruments that have not been developed yet.

Diagnostic clues that can determine whether or not comet material is primordial, and under what conditions it formed and has been subsequently processed, include the abundance of presolar grains, elemental abundance patterns, iron oxidation state, properties of refractory materials, isotopes of key elements and noble gases, and the presence of key hypervolatile species such as CO and $CO_2$. These clues are preserved in refractory silicate phases, salts, and ices of water and carbon dioxide in which hypervolatiles may be trapped. Furthermore, the basic properties of returned comet solids, including their texture, bulk mineralogy, and chemistry, will reveal the first ground truth for a comet nucleus, enabling direct comparison with Stardust samples of the coma dust of comet 81P/Wild-2, Interplanetary Dust Particles (IDPs), micrometeorites, meteorites, and OSIRIS-REx and Hayabusa2 samples. Comparisons of comet nucleus material to meteorites, and to IDPs in particular, will be critical for discerning whether cometary material represents a primordial fossil or if it has instead undergone a complex alteration history. If the interstellar model of comet origin is correct, IDP studies suggest a single aliquot of $< 10^{-6}$ g of cometary material may contain over 10 million presolar grains [75]. Comets may also contain a myriad of interstellar materials not yet found in asteroidal or IDP samples. If so, analysis of a returned sample from a comet nucleus will mark a new frontier in astrochemistry.

The origin (presolar or solar) of cometary water is unknown, but various processes in molecular clouds and protoplanetary disks can induce isotopic fractionations in both H and O [76-79] in water. As a result, these isotope systems are highly diagnostic for determining the origin (presolar or solar) of cometary water and the importance of comets as sources of Earth's water and other volatiles. Acquiring such data, however, has been impeded by the difficulty of accurately measuring isotopic ratios of comet volatiles by spacecraft and remote spectroscopy. For example, the D/H ratio of $H_2O$ in JFCs and Oort cloud comets range from 1 to ~3 times terrestrial [80;81] and have been shown to vary as a function of the comet's observed activity level, and possibly even along a comet's orbit. It may be that these isotopic variations are related to fractionation processes during sublimation from the comet or sampling different isotopic reservoirs in the nucleus. These inconsistent and variable isotopic measurements highlight the need for independent and precise measurements of H and O isotopes in cometary $H_2O$ to determine whether comets could have contributed a significant fraction of Earth's volatiles (inert gases and/or water). A direct sample of $H_2O$ ice from a comet nucleus would help to understand the causes of variable D/H ratio measurements of comet coma gas.

Comets and asteroids have delivered organic matter that may have supported the emergence of life on Earth [85]. Rosetta showed that ~45% of 67P's refractory material is organic by mass [24]. This implies that just one comet like 67P could have a mass of organic material (~4 Gtons) comparable to the biomass of Earth's oceans (~6 Gtons, [86].). However, not much is known about the prebiotic organic inventory of comets. A few prebiotic compounds have been observed remotely [34] but remote observations tell us little about complex molecules and refractories. Stardust showed the extraordinary potential of returned sample analysis for organics, where ultra-sensitive laboratory measurements were used to detect cometary organics in a micro-sample collected at 6.1 km/s [88]. Analysis of a returned nucleus sample will reveal for the first time the full ensemble of prebiotic organics in cometary materials. In particular, cometary materials that escaped aqueous alteration will contain unique inventories of prebiotic organics that did not survive in asteroids [89]. Finally, a major question in astrobiology is why terrestrial life almost exclusively uses left-handed amino acids and right-handed sugars. This homochirality has no known biochemical basis [90]. The chirality of returned comet samples will help to determine if comets contributed to homochirality in life on Earth.

A detailed understanding of the composition of comet nuclei, particularly their volatiles and organics, is also important to many other areas of planetary science. One example is the volatile deposits in permanently shadowed regions near the lunar poles. These deposits were predicted decades before their discovery [100-101], and strong evidence for their existence has been provided by several lunar missions [e.g., 102-103]. They have received attention in recent years because of their potential value as an in-situ source of $H_2O$ and $O_2$ for life support, and $H_2$ and $O_2$ for propellant [e.g., 104]. Comet impacts are likely to have been a significant contributor to these deposits, providing much of the $H_2O$, as well as other compounds that will act as contaminants for some purposes, but as important resources themselves (e.g., carbon and nitrogen) for others. Scientific understanding of these deposits and formulation of practical means for their extraction, purification, and use requires understanding of their cometary source materials.

The contributions of comets to the volatile inventories of Venus and Mars is another important problem. Noble gas data from SNC meteorites and understanding of the trapping of noble gases in ice has led to the suggestion that noble gases in the atmospheres of the terrestrial planets are dominated by a mixture of an internal component and a contribution from comets [107]. The notion that comets are a major source of atmospheric noble gases in the inner solar system is supported by recent Rosetta results [108]. Martian atmospheric C/N/$^{36}$Ar and $^{14}$N/$^{15}$N are consistent with late addition of cometary materials [108], though the Xe isotopic signature does not appear to require cometary input [106]. Comets may also have contributed volatiles to outer planet moons. Enceladus' D/H ratio suggests a cometary origin for its building blocks [105]. Large Kuiper Belt objects such as Pluto, Eris and (Neptune-captured) Triton all have nitrogen/methane surfaces along with other volatiles, reminiscent of the inventory on 67P.

4. **Strategic Value**

Comet Sample Return missions have been proposed for more than 30 years [96] and were recommended by both of the previous Planetary Science Decadal Surveys [35; 97]. The 2003 Decadal Survey (*New Frontiers in the Solar System: An Integrated Exploration Strategy*) recommended that NASA develop a medium-class mission to return a comet surface sample to Earth for laboratory analysis, and justified that recommendation with the following statement:

"No other class of objects can tell us as much as samples from a selected surface site on the nucleus of a comet can about the origin of the solar system and the early history of water and biogenic elements and compounds. Only a returned sample will permit the necessary elemental, isotopic, organic, and mineralogical measurements to be performed"

This recommendation was reaffirmed by the 2011 Decadal Survey (*Visions and Voyages*) and led to three proposals to the 2016 New Frontiers (NF) 4 Announcement of Opportunity (AO) [29-31], one of which was selected for Phase A study (CAESAR). The success of these proposals demonstrates that non-cryogenic comet sample return can be credibly proposed within a New Frontiers budget. These mission concepts are expected to be proposed to the coming 2022 NF5 AO. If not selected for NF5, we believe that non-cryogenic comet sample return should remain a high priority for the medium-class New Frontiers mission line in the decade 2023-2032.

While comet sample return missions, such as the cryogenic AMBITION concept, are discussed in the European Space Agency's (ESA) Voyage 2050 plan [92], the current planning for the ESA program indicates that large scale planetary missions (other than JUICE) are unlikely to be feasible programmatically or financially until at least the late 2030s [14], outside the scope of the current Planetary Science and Astrobiology Decadal Survey. Furthermore, a medium-class New Frontiers comet sample return mission will lay the groundwork and provide critical context for more complex cryogenic sample return concepts [33; 91-92].

## 5. Summary

Over the past several decades, substantial technology development has made non-cryogenic comet nucleus sample return possible within the New Frontiers program [29-31]. The results of ESA's Rosetta mission have demonstrated that comets are even more scientifically rich and enigmatic than previously thought, and validated that comet surface materials contain the refractory and volatile components necessary to address science questions regarding the nature of Solar System starting materials and how these fundamental components came together to form planets and give rise to life. The Philae lander, which bounced across the surface of 67P before coming to rest [98], has further provided constraints on the physical and chemical properties of comet surface materials and facilitated more robust engineering designs for comet nucleus sample acquisition. However, analyses of the non-volatile organic component of 67P, including molecules of high astrobiological interest such as amino acids, sugars, nucleobases, carboxylic acids, and associated isotopic and enantiomeric measurements, were not made by Rosetta or Philae and remain a wide open area of scientific research enabled by non-cryogenic comet sample return. Comet sample analyses can provide unparalleled knowledge about presolar history through the initial stages of planet formation to the origin of life. Considering the wealth of knowledge gleaned from less than a microgram of material returned from the coma of comet 81P/Wild 2 by Stardust [99], the return of ~100 g of non-volatile and volatile material from a comet nucleus will bring a leap forward in our understanding of Solar System history. In addition, sample return allows future researchers to address the fundamental science goals associated with comet nucleus sample return using ever more capable instruments and new scientific insights, providing an invaluable scientific resource for the global planetary science community for generations after the sample is acquired and returned to Earth.